\documentclass[a4paper]{jpconf}
\usepackage{amsmath}
\usepackage{graphicx}
\usepackage{latexsym}  
\usepackage{amsfonts}  
\usepackage{amssymb} 
\usepackage{color,mathtools}
\usepackage{subfig,cancel}
\usepackage{dsfont,xcolor}
\usepackage{url}
\usepackage[square,sort&compress,numbers]{natbib}
\usepackage{verbatim}
\usepackage[T1]{fontenc}
\usepackage[utf8x]{inputenc}

\newcommand{\be}{\begin{equation}}  
\newcommand{\ee}{\end{equation}}
\newcommand{\beq}{\begin{eqnarray}}  
\newcommand{\eeq}{\end{eqnarray}}

\newcommand{\adj}[2]{\hat{#1}_{#2}^{\dagger}}
\newcommand{\ope}[2]{\hat{#1}_{#2}^{\textcolor{white}{\dagger}}}

\newcommand{\del}[2]{\delta_{#1 #2}}

\begin{document}

\def\lipsum{Lorem ipsum dolor sit amet, consectetur adipiscing elit.
Morbi aliquet leo sed turpis imperdiet pellentesque. Phasellus vitae
urna elit, at laoreet lectus. Mauris pretium, leo hendrerit facilisis
consectetur, massa nisi vestibulum purus, vitae faucibus felis nunc
at leo. Integer posuere, risus sed hendrerit posuere, dui turpis
iaculis orci, sit amet volutpat nisl urna at tortor. Nunc sed aliquam
mi. Sed nibh mi, pulvinar ut commodo id, eleifend at sem. Aenean
ultricies tempus felis ac facilisis. Fusce viverra accumsan sodales.
Integer luctus nibh vitae massa fermentum non hendrerit quam
adipiscing. Integer sed elit nulla. Nulla facilisi. Nunc a nulla
nulla. Praesent.}
\def\impl{\ \Rightarrow \ }
\def\hc{\text{h.c.}}
\def\ud{{\mathrm{d}}}
\def\udl{\stackrel{\leftarrow}{\ud}}
\def\w{{\omega}}
\def\x{{\mathbf{x}}}
\def\y{{\mathbf{y}}}
\def\r{{\mathbf{r}}}
\def\k{{\mathbf{k}}}
\def\p{{\mathbf{p}}}
\def\s{\sigma}
\def\H{{\hat{H}}}
\def\T{{\hat{T}}}
\def\U{{\hat{U}}}
\def\V{{\hat{V}}}
\def\N{{\hat{N}}}
\def\Z{\mathcal{Z}}
\def\Htot{{\H_{\mathrm{tot}}}}
\def\Hch{{\H_{\mathrm{ch}}}}
\def\Hmol{{\H_{\mathrm{mol}}}}
\def\Heq{{\H_{\mathrm{eq}}}}
\def\uext{{U_{\mathrm{ext}}}}
\def\ubias{{U_{\mathrm{bias}}}}
\def\thyb{{t_{\mathrm{hyb}}}}
\def\eh{{\xi_{H}}}
\def\uhl{{U_{HL}}}
\def\un{{U_{0}}}
\def\im{{\mathrm{i}}}
\def\ex{{\mathrm{e}}}
\def\sgn{{\mathrm{sgn}}}
\def\co{{\mathcal{C}}}
\def\re{{\mathcal{R}}}
\def\twopi{{\frac{1}{2 \pi \im}}}
\def\cint{{\int_{\co}}}
\def\intg{\int_{\gamma}}
\def\coint{{\oint_{\co}}}
\def\ae{{\"a}}
\def\oe{{\"o}}
\def\unity{\mathds{1}}
\def\rhoh{{\hat{\rho}}}
\def\O{{\hat{O}}}
\def\A{{\hat{A}}}
\def\B{{\hat{B}}}
\def\C{{\hat{C}}}
\def\Cc{C}
\def\defi{\stackrel{\text{def.}}{=}}
\newcommand{\Id}{\mathbbm{1}}
\newcommand{\ket}[1]{| #1 \rangle}
\newcommand{\bra}[1]{\langle #1 |}
\newcommand{\ev}[2]{\langle #1 | #2 \rangle}
\newcommand{\mel}[3]{\langle #1 | #2 | #3 \rangle}
\newcommand{\od}[2]{\frac{\partial #1}{\partial #2}}
\newcommand{\der}[2]{\frac{\ud #1}{\ud #2}}
\newcommand{\nder}[3]{\frac{\ud^{#3} #1}{\ud #2^{#3}}}
\newcommand{\abs}[1]{\left| #1 \right|}
\newcommand{\com}[2]{\left[{#1} , {#2}\right]}
\newcommand{\acom}[2]{\left\{{#1} , {#2}\right\}}
\newcommand{\vpint}[2]{\mathcal{VP}\hspace{-5pt}\int_{#1}^{#2}}
\newcommand{\Tra}[1]{\mathrm{Tr}\{ #1 \}}
\newcommand{\Tc}[1]{\mathcal{T_{\g_{\text{K}}}} [ #1 ]}
\def\Tcc{\mathcal{T_{\g_{\text{K}}}}}
\def\Trr{\mathrm{Tr}}
\def\intx{{\int\hspace{-2pt}\ud\x\hspace{2pt}}}
\def\inty{{\int\hspace{-2pt}\ud\y\hspace{2pt}}}
\def\intp{{\int\hspace{-2pt}\ud\p\hspace{2pt}}}
\def\intxp{{\int\hspace{-2pt}\ud\x'\hspace{2pt}}}
\def\intyp{{\int\hspace{-2pt}\ud\y'\hspace{2pt}}}
\def\intw{{\int\hspace{-2pt}\frac{\ud \omega}{2\pi}}}
\def\intwp{{\int\hspace{-2pt}\frac{\ud \omega'}{2\pi}}}
\def\intwf{{\int\hspace{-2pt}\frac{\ud \omega}{2\pi}}f(\omega - \mu)}
\newcommand{\intn}[1]{\int\hspace{-2pt}\ud {#1} \hspace{2pt}}
\newcommand{\ihdt}[1]{\im\hbar\partial_{t_{#1}}}
\newcommand{\Gn}[4]{G_{(#1)}^{#2}(#3;#4)}
\newcommand{\Gnn}[3]{G^{#1}(#2;#3)}
\def\hbars{-\frac{\hbar^{2}}{2m}}
\def\deltas{\delta^{(3)}}
\renewcommand\Im{\operatorname{Im}}
\renewcommand\Re{\operatorname{Re}}

\def\tb{\bar{t}}  
\def\zb{\bar{z}}  
\def\tgb{\bar{\tau}}  
\def\bC{\mbox{\boldmath $C$}}  
\def\bG{\mbox{\boldmath $G$}}  
\def\bH{\mbox{\boldmath $h$}}  
\def\bK{\mbox{\boldmath $K$}}  
\def\bM{\mbox{\boldmath $M$}}  
\def\bN{\mbox{\boldmath $N$}}  
\def\bO{\mbox{\boldmath $O$}}  
\def\bQ{\mbox{\boldmath $Q$}}  
\def\bR{\mbox{\boldmath $R$}}  
\def\bS{\mbox{\boldmath $S$}}  
\def\bT{\mbox{\boldmath $T$}}  
\def\bU{\mbox{\boldmath $U$}}  
\def\bV{\mbox{\boldmath $V$}}  
\def\bZ{\mbox{\boldmath $Z$}}  
\def\unit{\mbox{\boldmath $1$}}
\def\bcalS{\mbox{\boldmath $\mathcal{S}$}}  
\def\bcalG{\mbox{\boldmath $\mathcal{G}$}}  
\def\bcalE{\mbox{\boldmath $\mathcal{E}$}}  
\def\bgG{\mbox{\boldmath $\Gamma$}}  
\def\bgL{\mbox{\boldmath $\Lambda$}}  
\def\bgS{\mbox{\boldmath $\mathit{\Sigma}$}}  
\def\itS{\mbox{$\mathit{\Sigma}$}}
\def\itG{\mbox{$\mathit{\Gamma}$}}
\def\Sigmaem{\itS_{\text{em}}}
\def\SemR{\itS_{\text{em}}^{\text{R}}}
\def\SemA{\itS_{\text{em}}^{\text{A}}}
\def\SR{\itS^{\text{R}}}
\def\SA{\itS^{\text{A}}}
\def\SM{\itS^{\text{M}}}
\def\bgr{\mbox{\boldmath $\rho$}}  
\def\a{\alpha}  
\def\b{\beta}  
\def\g{\gamma}  
\def\G{\Gamma}  
\def\d{\delta}  
\def\D{\Delta}
\def\GR{G^{\text{R}}}
\def\GA{G^{\text{A}}}
\def\GM{G^{\text{M}}}
\def\gR{g^{\text{R}}}
\def\gA{g^{\text{A}}}
\def\gM{g^{\text{M}}}
\def\SR{\itS^{\text{R}}}
\def\SA{\itS^{\text{A}}}
\def\SM{\itS^{\text{M}}}
\def\Gal{\itG_{\alpha}}
\def\Gbe{\itG_{\beta}}
\def\Val{V_\alpha}
\def\Valp{V_{\alpha'}}
\def\Vbe{V_\beta}

\def\eps{\epsilon}
\def\e{\mathbf{e}}

\def\ve{\varepsilon}  
\def\z{\zeta}  
\def\h{\eta}  
\def\th{\theta}  
\def\l{\lambda}  
\def\L{\Lambda}  

\def\n{\mathbf{n}}  
\def\np{\mathbf{n}'}
\def\kp{\mathbf{k}'}

\def\size1{\tiny}
\def\size2{\scriptsize}
\def\size3{\footnotesize}
\def\size4{\small}
\def\size5{\normalsize}
\def\size6{\large}
\def\size7{\Large}
\def\size8{\LARGE}
\def\size9{\huge}
\def\size10{\Huge}

\def\sp{\sigma '}  
\def\t{\tau}  
\def\f{\phi}  
\def\vf{\varphi}  
\def\F{\Phi}  
\def\c{\chi}  
\def\Q{\Psi}  
\def\q{\psi}  
\def\ua{\uparrow}  
\def\da{\downarrow}  
\def\de{\partial}  
\def\inf{\infty}  
\def\heff{h_{\text{eff}}}
\def\heffd{h_{\text{eff}}^{\dagger}}
\def\ra{\rightarrow}  
\def\grad{\mbox{\boldmath $\nabla$}}  
\def\ie{\emph{i.e.},}
\def\etal{\emph{et. al.}}
\def\eg{\emph{e.g.},}
\def\idt{\im\frac{\ud}{\ud t}}
\def\idz{\im\frac{\ud}{\ud z}}
\def\idzp{\im\frac{\ud}{\ud z'}}
\title{Time-dependent Landauer--Büttiker formula for transient
dynamics}

\author{Riku Tuovinen$^1$, Robert van Leeuwen$^{1,4}$, Enrico 
Perfetto$^{2}$ and Gianluca Stefanucci$^{2,3,4}$}

\address{$^1$ Department of Physics, Nanoscience Center, FIN 40014,
University of Jyv\"askyl\"a, Jyv\"askyl\"a, Finland}
\address{$^2$ Dipartimento di Fisica, Universit\`a di Roma Tor
Vergata, Via della Ricerca Scientifica 1, I-00133 Rome, Italy}
\address{$^3$ Laboratori Nazionali di Frascati, Istituto Nazionale di
Fisica Nucleare, Via E. Fermi 40, 00044 Frascati, Italy}
\address{$^4$ European Theoretical Spectroscopy Facility (ETSF)}

\ead{riku.m.tuovinen@jyu.fi}

\begin{abstract}
We solve analytically the Kadanoff--Baym equations for a 
noninteracting junction connected to an arbitrary number of 
noninteracting wide-band terminals. The initial equilibrium state is properly 
described by the addition of an imaginary track to the time contour. 
From the solution we obtain the time-dependent electron densities and 
currents within the junction. 
The final results are analytic expressions as a 
function of time, and therefore no time propagation is needed --- either in 
transient or in steady-state regimes. We further present and discuss some 
applications of the obtained formulae.           
\end{abstract}

\section{Introduction}

The Landauer--Büttiker formula \cite{landauer,buttiker} provides an
intuitive physical picture 
of the steady-state current flowing in a multi-terminal junction and
it is simple to implement. First one 
calculates the steady-state current $I_{\a\b}$ in terminal $\b$ 
carried by the scattering states originating from terminal $\a\neq\b$ 
and populated according to the electrochemical potential $\mu_\a$. 
Then one sums the difference $I_{\a\b}-I_{\b\a}$ between the currents 
flowing in and out terminal $\b$ over all terminals $\a\neq\b$. This 
gives the steady-state current $I_\b$ in terminal $\b$. 
% The currents 
% $I_{\a\b}$ are given by the integral with the Fermi function of 
% terminal $\a$ of the transmission probability for an electron to go 
% from terminal $\a$ to terminal $\b$.

The first microscopic derivation 
(based on the time-dependent Schrödinger equation) of the 
Landauer--Büttiker formula was given by Caroli and co-workers 
\cite{caroli1,caroli2}. They considered the terminals initially 
uncontacted and in equilibrium at different chemical potentials. Then 
they switched on the contacts and derived the Landauer--Büttiker 
formula as the long-time limit of the expectation value at time $t$ 
of the current operator. We will refer to this procedure as the {\em
partitioned approach}.

An alternative approach, more akin to the the way the experiments are
carried out, 
was proposed by Cini about a decade later \cite{cini}. He considered
the system initially contacted 
and in equilibrium at a unique chemical potential and then drove the
system out of equilibrium by 
applying a bias voltage between the terminals. We will refer to this
procedure
as the {\em partion-free approach}. In both approaches one recovers
the Landauer--Büttiker
formula due to the loss of memory of the initial preparation
\cite{stefanucci-rts}.

The microscopic derivation of the Landauer--Büttiker formula 
requires the evaluation of the expectation value 
$I_\b(t)=\langle\Psi(t)|\hat{I}_\b|\Psi(t)\rangle$ where 
$|\Psi(t)\rangle$ is the many-body state of the system at time $t$ 
and $\hat{I}_\b$ is the current operator. Since the electrons are 
noninteracting this expectation value can be rewritten as the sum 
over all occupied one-particle states $|\psi_k(t)\rangle =\ex^{-\im 
\hat{h} t}|\psi_k\rangle$ of 
$\langle\psi_k(t)|\hat{I}_\b|\psi_k(t)\rangle$. Here $\hat{h}$ is the 
Hamiltonian of the contacted and biased system whereas 
$|\psi_k\rangle$ are the eigenstates of the Hamiltonian $\hat{h}_0$
which describes
either the non-biased uncontacted system (in the partitioned approach) or the
non-biased contacted system 
(in the partition-free approach). For the evaluation of
$\langle\psi_k(t)|\hat{I}_\b|\psi_k(t)\rangle$ 
one could naively insert a complete set of eigenstates
$|\phi_q\rangle$ of $\hat{h}$ 
and evaluate the overlaps $\langle\psi_k|\phi_q\rangle$. This
procedure is, however,
numerically lengthy and unstable due to the singular $\delta$-like
contribution to the 
overlaps. The calculation of $I_\b(t)$ is most easily carried out
using nonequilibrium 
Green's functions \cite{meir-wingreen,jauho}. This mathematical tool
when applied to quantum transport in multi-terminal junctions
provides a natural framework to calculate the current at all times
and not only at the steady state.

In fact, there have been several
attempts to generalize the Landauer--Büttiker formula to the time
domain. Here we mention the work of Pastawski who derived a formula
for $I_\b(t)$ using the partitioned approach in the linear response
and adiabatic regime \cite{pastawski}. An important step forward in
the calculation of $I_\b(t)$ was done by Jauho et al. \cite{jauho}.
These authors used the partitioned approach to write $I_\b(t)$ as a
double integral (over time and energy) of the trace over the junction
degrees of freedom of a calculable combination of Green's functions
in the same region. In the special case of terminals with a wide band
and of junctions with one single level it is possible to perform the
time-integral and obtain a time-dependent version of the
Landauer--Büttiker formula. This formula was then derived in Ref.
\cite{stefanucci-rts} using the partition-free approach, thus
confirming the loss of memory of the initial preparation.

The derivation of a time-dependent Landauer--Büttiker formula for
arbitrary junctions would be extremely useful to interpret the
oscillations and damping times typically observed in the transient
current after the sudden switch on of a bias. A progress in this
direction was done in Ref. \cite{perfetto} where the authors derived
a time-dependent Landauer--Büttiker formula for the spin current of
a single-level junction.

In this work we generalize the results of Ref. \cite{perfetto} to
junctions of any shape and dimensions using the wide-band limit
approximation (WBLA) for noninteracting electrons (Secs. \ref{sec:setup} and \ref{sec:deriv}). 
Furthermore we also derive a general formula for the time-dependent
one-particle density matrix which can be used to calculate the local
density and current density. We will work in
the partition-free approach which is conceptually easier since it
does not involve the subtle issue of different chemical potentials in
equilibrium. The final formulae for the current and the one-particle
density matrix have the merit of elucidating the relative importance
of the electronic transitions at a certain time. As an illustration
we will use these formulae to calculate the transient response of a
ring-shaped junction (Sec. \ref{sec:results}).

\section{Assumptions and set-up}\label{sec:setup}
We investigate the following quantum transport setup: An arbitrary
number of metallic leads ($\a$) acting as charge-carrier reservoirs
are connected to a lattice network acting as a molecular device
($C$). We assume that the electron transport is ballistic and 
therefore neglect the electron--electron interactions. We will also assume 
that the energy
eigenvalues of the Hamiltonian of the molecular device are well inside the
continuous energy spectrum of the leads and use the WBLA. 

The described set-up is characterized by the following
Hamiltonian:
\be\label{eq:nonintham}
\H =
\sum_{k\a,\sigma}\eps_{k\a}^{\textcolor{white}{\dagger}}\ope{n}{k\a,\sigma}
+
\sum_{mn,\sigma}T_{mn}^{\textcolor{white}{\dagger}}\adj{d}{m,\sigma}\ope{d}{n,\sigma}
+
\sum_{mk\a,\sigma}\left[T_{mk\a}^{\textcolor{white}{\dagger}}\adj{d}{m,\sigma}\ope{d}{k\a,\sigma}+T_{k\a
m}^{\textcolor{white}{\dagger}}\adj{d}{k\a,\sigma}\ope{d}{m,\sigma}\right]
\ .
\ee
The first term accounts for the leads with $k\a$ indexing the $k$:th
basis function of the $\a=1,2,3,\ldots$ lead. The single-particle spectrum of
the leads is $\eps_{k\a}$ and the number operator in the leads is
expressed in terms of the creation and annihilation
operators as $\ope{n}{k\a,\s}=\adj{d}{k\a,\s}\ope{d}{k\a,\s}$, with $\s$  the spin index. The
second term is for the molecular device, or central region, (indices $m$ and $n$) 
with creation and
annihilation operators $\adj{d}{m,\s}$ and $\ope{d}{m,\s}$ and 
hoppings $T_{mn}$  between sites $m$ and $n$. The last
term is for the coupling between the central region and the leads
with hoppings $T_{mk\a}$. 

At times $t<t_0$ the system is in thermal equilibrium at inverse
temperature $\b$ and chemical potential $\mu$, the density matrix
having the form $\hat{\rho} = \frac{1}{\Z}\ex^{-\b(\H -\mu\N)}$ where
$\Z$ is the grand-canonical partition function. At $t=t_0$ the
lead energy levels are suddenly shifted by some constant value,
$\eps_{k\a} \to \eps_{k\a}+V_\a$, to model the sudden switch-on of an external bias
voltage in the $\a$:th lead. This means that the system is driven
out of equilibrium and charge carriers start to flow through the
central region. To calculate the time-dependent current we use 
the equations of motion for the one-particle Green's 
function on the Keldysh contour $\g_{\text{K}}$. This quantity is defined as the ensemble average
of the contour-ordered product of particle creation and
annihilation operators in the Heisenberg picture
\be\label{eq:greenf}
G_{rs}(z,z') =
-\im\langle\Tc{\ope{d}{r,\mathrm{H}}(z)\adj{d}{s,\mathrm{H}}(z')}\rangle
\ee
where the indices $r$, $s$ can be either indices in the leads or in 
the central region and the
variables $z$, $z'$ run on the contour\footnote{The
contour has a forward and a backward branch on the real-time axis,
$[t_0,\infty[$, and also a vertical branch on the imaginary axis,
$[t_0,t_0-\im\b]$ with inverse temperature $\b$, see e.g.
\cite{petriprb}.}. The matrix $\bG$ with matrix elements 
$G_{rs}$ satisfies the equations of motion
\beq
\left[\idz - \bH(z)\right]\bG(z,z') & = & \delta(z,z')\unit
\label{eq:eom-left} \ , \\
\bG(z,z')\left[-\im \frac{\udl}{\ud z'} - \bH(z')\right] & = &
\delta(z,z')\unit \label{eq:eom-right} \ ,
\eeq
with Kubo--Martin--Schwinger (KMS) boundary conditions. Here  
$\bH(z)$ is the single-particle Hamiltonian.
In the basis $k\a$ and $m$ the matrix $\bH$ has
the following block structure
\be\label{eq:ham-matrix}
\bH = \begin{pmatrix}h_{1 1} & 0 & 0 & \cdots & h_{1 \Cc} \\
                   0 & h_{2 2} & 0 & \cdots & h_{2 \Cc} \\
                   0 & 0 & h_{3 3} & \cdots & h_{3 \Cc} \\
                   \vdots & \vdots & \vdots & \ddots & \vdots \\
                   h_{\Cc 1} & h_{\Cc 2} & h_{\Cc 3} & \cdots &
h_{\Cc \Cc}
    \end{pmatrix} \ ,
\ee
where $(h_{\a \a'})_{kk'} = \del{\a}{\a'}\del{k}{k'}\eps_{k\a}$
corresponds to the leads, $(h_{\a \Cc})_{km} = T_{k\a m}$ is the
coupling part, and $(h_{\Cc \Cc})_{mn} = T_{mn}$ accounts for the
central region. As the system is initially in thermal equilibrium we 
have that for $z$ on the vertical track of the contour
$\eps_{k\a}(z)=\eps_{k\a}-\mu$, $T_{k\a m}(z)=T_{k\a m}$ and 
$T_{mn}(z)=T_{mn}-\mu\d_{mn}$. On the other hand for $z$ on the 
horizontal branches we have $\eps_{k\a}(z)=\eps_{k\a}+V_{\a}$, $T_{k\a m}(z)=T_{k\a m}$ and 
$T_{mn}(z)=T_{mn}$. Due to the coupling between the central region 
and the leads the matrix $\bG$ has nonvanishing entries 
everywhere
\be\label{eq:green-self-matrix}
\bG = \begin{pmatrix} G_{11} & \cdots & G_{1\Cc} \\
                    \vdots & \ddots & \vdots \\
                    G_{\Cc1} & \cdots & G_{\Cc\Cc}
    \end{pmatrix} .
\ee
In the next Section we solve the equations of motion \eqref{eq:eom-left}
and~\eqref{eq:eom-right} for the Green's function $G_{CC}$ projected 
in the central region.

\section{Derivation of the time-dependent density and
current}\label{sec:deriv}

\subsection{Projecting the equation of motion}
We project the equation of motion \eqref{eq:eom-left} onto regions $\Cc\Cc$ and $\a \Cc$. The
equation for $G_{\a C}$ can be integrated using  the Green's function $g_{\a \a}(z,z')$ of the isolated
$\a$:th reservoir. This Green's function solves the equation of
motion $\left[\idz-h_{\a \a}\right]g_{\a \a}(z,z') = \delta(z,z')$ as 
well as the adjoint equation with KMS boundary conditions.
Introducing the embedding self-energy (with indices in region $C$)
\be
\Sigmaem(z,z') = 
\sum_{\a}\Sigma_{\a}(z,z')\quad;\quad\Sigma_{\a}(z,z')=
h_{\Cc\a}g_{\a \a}(z,z')h_{\a
\Cc}
\label{sigembdef}
\ee
we obtain the equation of motion for the Green's function
of the central region (the subscripts $\Cc\Cc$ are
omitted from now on)
\be
\left[\idz-h\right]G(z,z')  =  \delta(z,z') +
\int_{\g_{\text{K}}}\!\!\!\ud\zb \Sigmaem(z,\zb)G(\zb,z')
\label{eq:nonint-eom-left}
\ee
The adjoint equation of motion can be derived 
similarly and read~\cite{petriprb}
\be
G(z,z')\left[-\im\frac{\udl}{\ud z'}-h\right]  =  \delta(z,z') +
\int_{\g_{\text{K}}}\!\!\!\ud\zb G(z,\zb)\Sigmaem(\zb,z') \ .
\label{eq:nonint-eom-right}
\ee
The embedded 
equations of motion for $G$ have the same structure as the 
Kadanoff--Baym equations (KBE), the difference being that the many-body self-energy is 
replaced by the embedding self-energy. In the case of interacting
electrons with an interaction only in the central region  Eqs. 
(\ref{eq:nonint-eom-left}) and (\ref{eq:nonint-eom-right}) are 
modified by the addition of the many-body self-energy $\itS$ to the 
embedding self-energy $\Sigmaem$, i.e., $\Sigmaem\ra\Sigmaem+\itS$. Since $\itS=\itS[G]$  is a functional of the 
Green's function in  region $C$ the  embedded 
equations of motion in the interacting case constitute  a closed set of integro-differential, 
nonlinear equations for $G$~\cite{petriprb}. 
The simplification brought by the absence of interactions is that the 
KBE (\ref{eq:nonint-eom-left}) and (\ref{eq:nonint-eom-right}) are linear in $G$ since  
the embedding self-energy is completely specified by the parameters 
of the Hamiltonian. 

The density and current density can be extracted from the lesser 
component of the Green's function at equal time. We denote by 
$z=t_{-}$ the contour point on the forward branch, $z=t_{+}$ the 
contour point on the backward branch and $z=t_{0}-\im\t$ the 
contour point on the vertical track. The Keldysh components
lesser ($<$), greater ($>$), retarded (R), advanced (A), left 
($\lceil$), right ($\rceil$) and Matsubara (M) of a function 
$k(z,z')$ on the contour are defined according to~\cite{svlbook1}
\beq
k^{<}(t,t')&=&k(t_{-},t'_{+}) \\
k^{>}(t,t')&=&k(t_{+},t'_{-}) \\
k^{\rm R}(t,t')&=&+\theta(t-t')\left[k^{>}(t,t')-k^{<}(t,t')\right] \\
k^{\rm A}(t,t')&=&-\theta(t'-t)\left[k^{>}(t,t')-k^{<}(t,t')\right] \\
k^{\lceil}(\t,t')&=&k(t_{0}-\im\t,t')\\
k^{\rceil}(t,\t)&=&k(t,t_{0}-\im\t)\\
k^{\rm M}(\t,\t')&=&k(t_{0}-\im\t,t_{0}-\im\t')
\eeq
To generate an equation for $G^{<}$ 
we subtract Eq.~\eqref{eq:nonint-eom-right} from
Eq.~\eqref{eq:nonint-eom-left} and set $z=t_-$, $z'=t_+'$. Taking 
into account that $h(z)=h$ is independent of $z$ for $z$ on the horizontal 
branches we get at equal time
\beq\label{eq:deriv3}
\idt G^<(t,t) - \com{h}{G^<(t,t)} & = & \left[\SemR \cdot G^< - \GR
\cdot \Sigmaem^< + \Sigmaem^< \cdot \GA - G^< \cdot \SemA
\right](t,t) \nonumber \\
& + & \left[\Sigmaem^{\rceil}\star G^{\lceil} - G^{\rceil}\star
\Sigmaem^{\lceil}\right](t,t) \ ,
\eeq
where we defined $\left[f \cdot g\right](t,t') =
\int_{t_0}^{\infty}\ud \tb f(t,\tb)g(\tb,t')$ and $\left[f \star
g\right](t,t') = -\im\int_{0}^{\b}\ud \t f(t,\t)g(\t,t')$. 
Equation (\ref{eq:deriv3}) can also be written as
\be\label{eq:final-glss}
\idt G^<(t,t)-\com{h}{G^<(t,t)} =
-\left[\GR\cdot\Sigmaem^<+G^<\cdot\Sigmaem^{\text{A}}+G^{\rceil}\star\Sigmaem^{\lceil}\right](t,t)+\hc
\ee
where we used the properties of $G$ and $\Sigmaem$ under complex 
conjugation~\cite{svlbook1}.

Let us comment Eq.~\eqref{eq:final-glss} briefly.
Setting the right-hand side to zero we see that Eq.~\eqref{eq:final-glss} reduces 
to the Liouville equation for the one-particle density matrix $\rho=-\im G$ 
of the isolated central region. Thus the embedding self-energy 
accounts for the openness of region $C$.
The first term inside the square brackets is a convolution between the
propagator in region $C$, $\GR$, and $\Sigmaem^<$. Since 
$\Sigmaem^<$ is proportional to the probability of finding an electron in 
the leads this term can be interpreted as a source term, i.e., a term 
that 
describes the injection of electrons into region $C$.
The second term has the opposite structure: a
propagator in the leads, $\Sigmaem^{\rm A}$, is convoluted with $G^<$ which is 
propotional to the probability of finding an electron in region $C$. 
Thus this term can be interpreted as a drain term and
is responsible for damping and equilibration effects.
The last term inside the square
brackets accounts for the initial preparation of the system.
In the partioned approach this term would be zero since the hopping 
integrals $T_{k\a m}=0$ in equilibrium. However, in the 
partition-free approach this term is nonzero and accounts for 
the initial coupling of the central region to the leads.

More generally convolutions along the vertical track carry information 
on the initial preparation of the system. For instance for a system of 
interacting electrons we can  either start with a noninteracting 
system and then switch on the interaction in real time or we can 
start with a 
system already interacting. In the latter case the many-body 
self-energy is nonvanishing on the vertical track and the convolution 
$G\star\itS$ accounts for the effects of initial correlations.

\subsection{Self-energy and Green's function calculations}

The solution of Eq.~\eqref{eq:final-glss} requires first to calculate the
Matsubara component $\GM$, and then from $\GM$ the right and left
component $G^\rceil$ and $G^\lceil$. The Matsubara component $\GM$ 
can be determined from the retarded/advanced components by analytic 
continuation, see below. Since the equations for $\GM$, $G^\rceil$ and $G^\lceil$ 
contain the embedding self-energy the preliminary step is to obtain an  
expression for  $\Sigmaem$.

Having a time-independent Hamiltonian (on the horizontal branches of
the contour) the retarded/advanced components of the
self-energy depend only on the time difference
\be\label{eq:adv-self}
\SA_{\a,mn}(t,t') = \intw \ex^{-\im \w (t-t')}\sum_k
T_{m k\a}\,g_{k\a}^{\mathrm{A}}(\w)\, T_{k\a n}
\ee
where $g_{k\a}$ is the diagonal element of the 
Green's function $g_{\a\a}$ of the isolated $\a$:th lead, see Eq. 
(\ref{sigembdef}). The retarded component
of the self-energy is found by conjugating
$\SR_{\a}(t,t')=\left[\SA_{\a}(t',t)\right]^\dagger$. According to the
WBLA the Fourier transform of $\SA_{\a}$ is frequency independent
\be\label{eq:wbla}
\SA_{\a,mn}(\w) =\sum_k
T_{m k\a}\,g_{k\a}^{\mathrm{A}}(\w)\, T_{k\a n}= \sum_k
T_{m k\a}\frac{1}{\w-\eps_{k\a}-V_\a-\im\eta} T_{k\a n} =
\frac{\im}{2}\itG_{\a,mn} \ 
\ee
which implies that $\SA_{\a}$ is also independent of the external bias voltage
$V_\a$. The time-dependent self-energy of Eq. (\ref{eq:adv-self})
is therefore
\be\label{eq:self-wbla}
\SA_{\a,mn}(t,t') = \intw \ex^{-\im\w(t-t')}\SA_{\a,mn}(\w) =
\frac{\im}{2}\itG_{\a,mn}\delta(t-t') \ .
\ee
Within the WBLA we can easily calculate the two other self-energy 
components in
Eq.~\eqref{eq:final-glss} (see \ref{sec:app-self})
\beq
\itS_{\a,mn}^{\lceil}(\t,t) & = &
\itG_{\a,mn}\frac{1}{-\im\b}\sum_{q}\ex^{-\w_q\t}\intw\frac{\ex^{\im(\w+\Val)t}}{\w_q-\w+\mu}
\ , \\
\itS_{\a,mn}^<(t,t') & = &
\im\itG_{\a,mn}\intwf\ex^{-\im(\w+\Val)(t-t')} \ ,
\eeq
where the sum over $q$ is a sum over the Matsubara frequencies $\w_q= 
\frac{(2q+1)\pi}{-\im\b}$, and the function $f$ is the Fermi function, $f(\w) = 1/[\ex^{\b\w}+1]$.

Having the explicit form of the self-energy components
we can derive expressions
for the Green's function. For the following calculations it is
convenient to define the \emph{effective} Hamiltonian $\heff =
h-\frac{\im}{2}\itG \impl \heffd = h+\frac{\im}{2}\itG$, where $\itG
= \sum_{\a}\itG_\a$. This effective Hamiltonian is therefore
non-hermitean. The two Green's function components in the square
brackets of Eq.~\eqref{eq:final-glss} read (see
\ref{sec:app-green}) 
\beq
G^\rceil(t,\t) & = & \ex^{-\im\heff t}\Big[\GM(0,\t)-\int_0^t\ud
t'\ex^{\im\heff
t'}\int_0^\beta\ud\tgb\itS_{\rm em}^\rceil(t',\tgb)\GM(\tgb,\t)\Big] \ , \\
\GR(t,t') & = & -\im\theta(t-t')\ex^{-\im\heff(t-t')} \ ,
\label{gret}
\eeq
with $\GM$ the Matsubara Green's function. $\GM$ can be obtained from 
$\GR$ and $\GA$ by analytic continuation since 
$\GM(\w_{q})=G^{\rm R}(\w_{q}+\mu)$ if ${\rm Im}[\w_{q}]>0$ and 
$\GM(\w_{q})=\GA(\w_{q}+\mu)$ if ${\rm Im}[\w_{q}]<0$, see 
\ref{sec:app-green}.

Now we have all ingredients to calculate the convolutions in Eq.~\eqref{eq:final-glss}. 
We report here the final results and refer to \ref{sec:app-terms} for
details. The three terms read
\beq
\left[\GR\cdot\itS_{\rm em}^<\right](t,t) & = &
\im\sum_\a\intwf\left[1-\ex^{\im(\w+\Val-\heff)}\right]\GR(\w+\Val)\itG_\a
\ ,\label{eq:first-term} \\
\left[G^<\cdot\SA_{\rm em}\right](t,t) & = & \frac{\im}{2}G^<(t,t)\itG \ ,
\label{eq:second-term} \\
\left[G^\rceil\star\itS_{\rm em}^\lceil\right](t,t) & = &
\im\intwf\sum_\a\ex^{\im(\w+\Val-\heff)t}\GR(\w)\itG_\a \ .
\label{eq:third-term}
\eeq

\subsection{Solving Eq.~\eqref{eq:final-glss} for $G^<(t,t)$}
We  insert
Eqs.~\eqref{eq:first-term},~\eqref{eq:second-term}
and~\eqref{eq:third-term} into Eq.~\eqref{eq:final-glss} and get
\beq\label{eq:glss-deriv1}
& & \idt G^<(t,t) - \big[h,G^<(t,t)\big] \nonumber \\
& = &
-\left\{\im\sum_\a\intwf\left[1-\ex^{\im(\w+\Val-\heff)t}\right]\GR(\w+\Val)\itG_\a
+\frac{\im}{2}\itG G^<(t,t) \right. \nonumber \\
& & \hspace{10pt} + \left. \im\intwf
\sum_\a\ex^{\im(\w+\Val-\heff)t}\GR(\w)\itG_\a \right\} + \hc 
\eeq
This is a nonhomogeneous, linear, first-order 
differential equation for $G^{<}(t,t)$ and, therefore, can be solved 
explicitly. The solution is worked out in \ref{sec:app-manip} and reads 
\beq\label{eq:glss-final2}
-\im G^{<}(t,t) & = & \intwf \sum_{\a} \Big\{ \ A_{\a}(\w +
\Val)\nonumber \\
& + & \Val\left[\ex^{\im(\w + \Val-\heff)t} \GR(\w)A_\a(\w + \Val) +
\hc\right] \nonumber \\
& + & \Val^2\ex^{-\im\heff t}\GR(\w) A_\a(\w +
\Val)\GA(\w)\ex^{\im\heffd t} \ \Big\} \ ,
\eeq
where we introduced the partial spectral function as
\be\label{eq:spectral-function}
A_\a(\w) =  \GR(\w)\Gal\GA(\w) \ .
\ee
The full nonequilibrium spectral function is $A(\w)=\sum_{\a}A_{\a}(\w)$.

Given the original complexity of the problem the final result
is surprisingly compact.
Equation \eqref{eq:glss-final2} is an explicit closed formula for the 
equal-time $G^{<}$ or, equivalently, for the reduced one-particle 
density matrix. All the
quantities inside the integral can be calculated separately, and no
time-propagation nor self-consistency algorithms are needed. Also, we
may extract several physical properties:
\begin{enumerate}
\item With no external bias, $\Val=0$, only the first row 
contributes. This term correctly gives the equilibrium value of the 
equal-time $G^{<}$ since at zero bias $\sum_{\a}A_{\a}(\w)$ is the 
equilibrium spectral function.
\item Both the second and the third row vanish exponentially in the
long-time limit, and the equal-time $G^{<}$ approaches a unique 
steady-state value.
\item The transient dynamics is given by the second and the third row. By
inserting a complete set of eigenstates of the effective Hamiltonian
$\heff$ we notice that:
\begin{enumerate}
\item The second row gives rise to oscillations with frequency $\w_j
= |\mu+\Val-\eps_j^{\text{eff}}|$ where $\eps_j^{\text{eff}}$ is the 
real part of the $j$:th complex eigenvalue of $\heff$. 
These oscillations correspond to transitions
between the biased Fermi level of the leads and the resonant levels
of the central molecule.
\item The third term accounts for intramolecular transitions and 
leads to oscillations with
frequency $\w_{jk} = |\eps_j^{\text{eff}}-\eps_k^{\text{eff}}|$.
These oscillations are visible only if the effective Hamiltonian
$\heff$ does \emph{not} commute with $\itG_\a$. In the case that 
$[\heff,\itG_{\a}]=0$ the time dependence of the
third term is  of the form $\ex^{-\im\heff t
+ \im\heffd t} = \ex^{-\itG t}$.
\end{enumerate}
\end{enumerate}

\subsection{Current calculation}
The time-dependent current through the interface between the central
region and the $\a$:th reservoir is calculated from the following
equation~\cite{petriprb}
\be\label{eq:current}
I_\a(t) =
4q\Re\left\{\Trr\left[\itS_{\a}^<\cdot\GA+\itS_{\a}^{\text{R}}\cdot
G^<+\itS_{\a}^\rceil\star G^\lceil\right](t,t)\right\}
 \ .
\ee
For the terms inside Eq.~\eqref{eq:current} we proceed in the same 
manner as we did previously to obtain the  results in Eqs.~\eqref{eq:first-term},
~\eqref{eq:second-term} and~\eqref{eq:third-term}:
\beq
\left[\itS_{\a}^<\cdot \GA\right](t,t) & = &
\im\intwf\itG_\a\GA(\w+\Val)\left[1-\ex^{-\im(\w+\Val-\heffd)t}\right]
\ , \label{eq:first-term-current} \\
\left[\itS_{\a}^{\text{R}}\cdot G^<\right](t,t) & = &
-\frac{\im}{2}\itG_\a G^<(t,t) \ , \label{eq:second-term-current} \\
\left[\itS_{\a}^\rceil \star G^\lceil\right](t,t) & = &
\im\intwf\itG_\a\GA(\w)\ex^{-\im(\w+\Val-\heffd)t} \ .
\label{eq:third-term-current}
\eeq
Inserting these results into Eq.~\eqref{eq:current} and taking into 
account the explicit expression for $G^<(t,t)$ in Eq. 
(\ref{eq:glss-final2}) we get
\beq\label{eq:current-deriv1}
I_\a(t) & = & -2 \intwf \sum_{\b} \Trr \Big\{ \nonumber \\
&   & \Gal \GR(\w + \Vbe) \Gbe \GA(\w + \Vbe) - \Gal \GR(\w + \Val)
\Gbe \GA(\w + \Val) \nonumber \\
& + & \Vbe\left[\Gal\ex^{\im(\w + \Vbe -
\heff)t}\GR(\w)\left(-\im\del{\a}{\b}\GR(\w + \Vbe) + A_\b(\w +
\Vbe)\right) + \hc \right]\nonumber \\
& + & \Vbe^2\Gal\ex^{-\im\heff t}\GR(\w)A_\b(\w +
\Vbe)\GA(\w)\ex^{\im\heffd t}\ \Big\} \ .
\eeq
The physical interpretation of the terms in Eq.~\eqref{eq:current-deriv1}
is similar to the one after Eq.~\eqref{eq:glss-final2}.
We have a steady-state part given by the first row, and
a time-dependent part given by the second and the third rows. The 
time-dependent part
vanishes exponentially in the long-time limit and the oscillations in
the current have the same structure as in the reduced one-particle 
density matrix.

\section{Results}\label{sec:results}
\begin{figure}[tbp]
\begin{minipage}{\textwidth}
\centering
\includegraphics[width=0.85\textwidth]{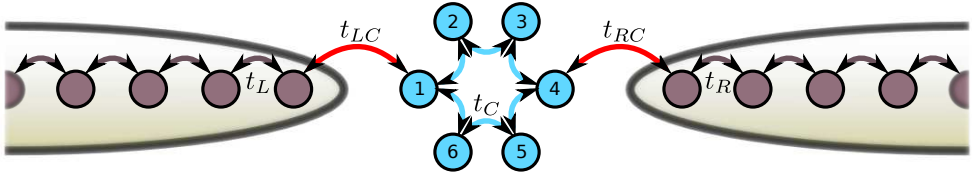}
\caption{\label{fig:schematic-sites}Six-site ring coupled
symmetrically to one-dimensional TB leads.}
\end{minipage}\hspace{2pc}%
\end{figure}

\begin{figure}[tbp]
\begin{minipage}{\textwidth}
\centering
\includegraphics[width=0.85\textwidth]{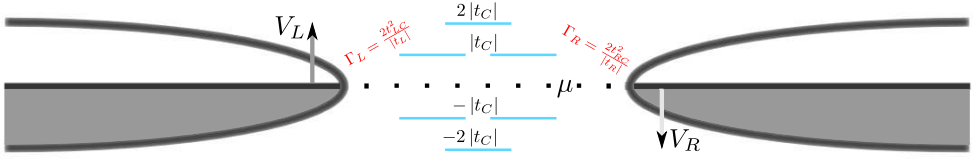}
\caption{\label{fig:schematic-levels}Schematic of the transport setup through
                                     the energy levels of the molecule.}
\end{minipage}\hspace{2pc}%
\end{figure}

Let us consider a
six-site tight-binding ring connected to two tight-binding,
semi-infinite, one-dimensional leads as shown schematically in
Figs.~\ref{fig:schematic-sites} and~\ref{fig:schematic-levels}.
The parameters according to the notation in the figures are $t_C = -2.0$
(hopping in the molecule), chemical potential $\mu=0$ and zero 
temperature ($\beta\to\infty$). We choose the hopping $t_L = t_R$ in the left/right lead to be much 
larger than any other energy scale. Then $\Sigma_{\alpha}^{\rm A}(\w\sim 
\mu)=\im\, t^{2}_{\a C}/t_{\alpha} +O(1/t_{\a}^{2})$ where $t_{\alpha C}$ is the 
hopping between the molecule and the leads. For this situation the 
WBLA with $\Gal=2t^{2}_{\alpha C}/t_{\alpha}$ is a very good approximation. 
We study the weak coupling case $\G_{L,R} = 0.1$ and drive the system 
out of equilibrium by the sudden switch-on of a bias  $V_L=V=-V_R$.
We analyze the contribution of different terms in the charge current 
corresponding to different physical features as discussed below 
Eq.~\eqref{eq:glss-final2}. In Eq.
\eqref{eq:current-deriv1} the first row is 'steady state', the
second row consists of '1st term, a' and '1st term, b' and the third row is '2nd term'.
The second row is divided into two parts [$\sim\GR(\w)\GR(\w+\Vbe)$ and 
$\sim\GR(\w)A_\b(\w+\Vbe)$] since they give rise to different features.

\begin{figure}[h]
\begin{minipage}{0.45\textwidth}
\centering
\includegraphics[width=\textwidth]{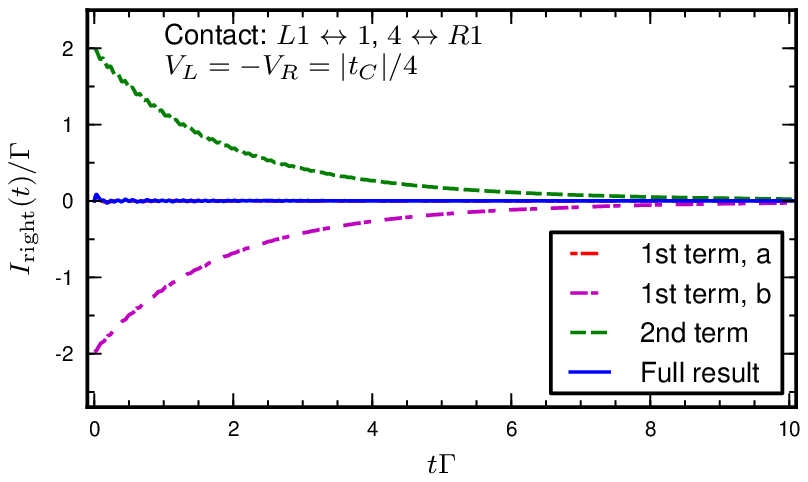}
\caption{\label{fig:res1}Different terms of the time-dependent current in units of $\G$
                         through the right interface with symmetric coupling and weak bias.}
\end{minipage}\hspace{2pc}%
\begin{minipage}{0.45\textwidth}
\centering
\includegraphics[width=\textwidth]{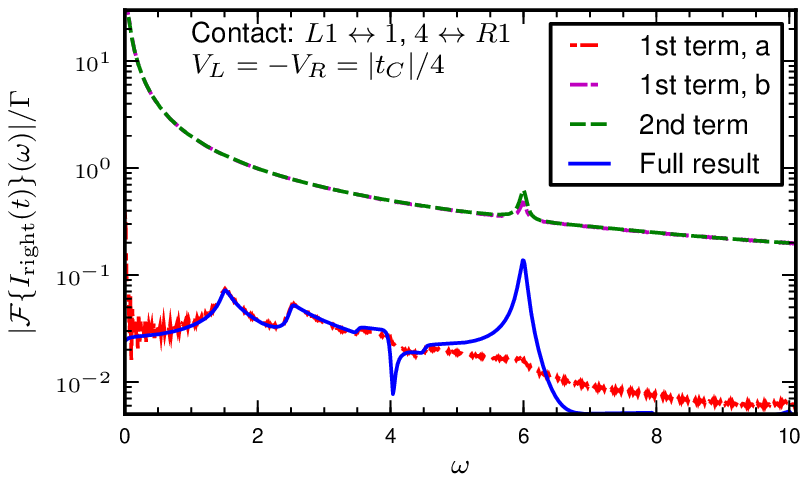}
\caption{\label{fig:res2}Absolute value of the Fourier transform of 
the terms  of the current  in units of $\G$. \\}
\end{minipage}
\end{figure}

In Figs.~\ref{fig:res1} and~\ref{fig:res2} we plot the current 
through the right interface and see that weakly biased 
leads, $V = 0.5$, give a negligible steady-state current. Transitions between the biased leads and
the molecule are captured by the '1st term, a'. This is confirmed by 
the peak in the Fourier
spectrum at $\w_j = \eps_j^{\text{eff}} \pm V$. Transitions between the molecular
levels are accounted for by the '2nd term', as it can be seen in the 
Fourier transform with a peak at $\w=6$. In addition
to our previous observations: (1) the '1st term, b' also gives rise to intramolecular
transitions and (2) there seems to be no intramolecular transitions at $\w=2$
or $\w=4$. By expanding Eq. (\ref{eq:current-deriv1}) in the eigenbasis of the
effective Hamiltonian and manipulating the terms further one can show that the 
the '1st term, b' contains a term of the form $\ex^{-\im(\eps_j^{\text{eff}} - \eps_k^{\text{eff}})t}$
which explains the first finding. The second finding suggests that there is some 
underlying selection rule for some of the energy levels and hence 
that some levels do not participate to the transport process.

\begin{figure}[h]
\begin{minipage}{0.45\textwidth}
\centering
\includegraphics[width=\textwidth]{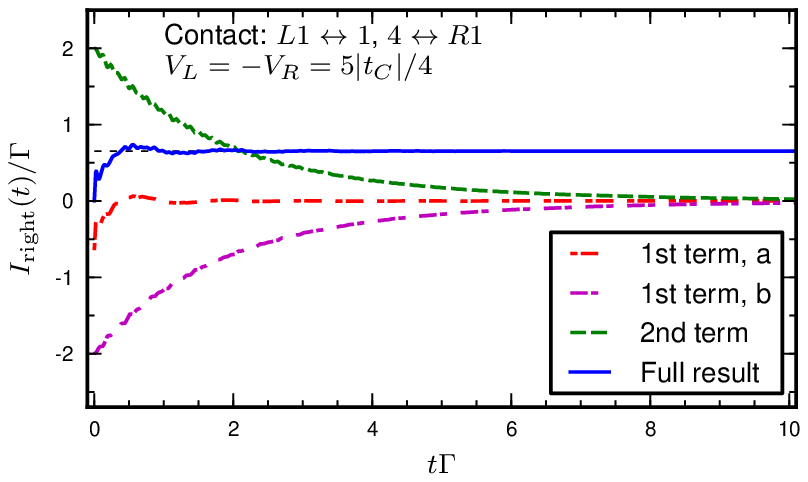}
\caption{\label{fig:res1a}Different terms of the time-dependent current in units of $\G$ through
                         the right interface with symmetric coupling and strong bias. 
                         }
\end{minipage}\hspace{2pc}%
\begin{minipage}{0.45\textwidth}
\centering
\includegraphics[width=\textwidth]{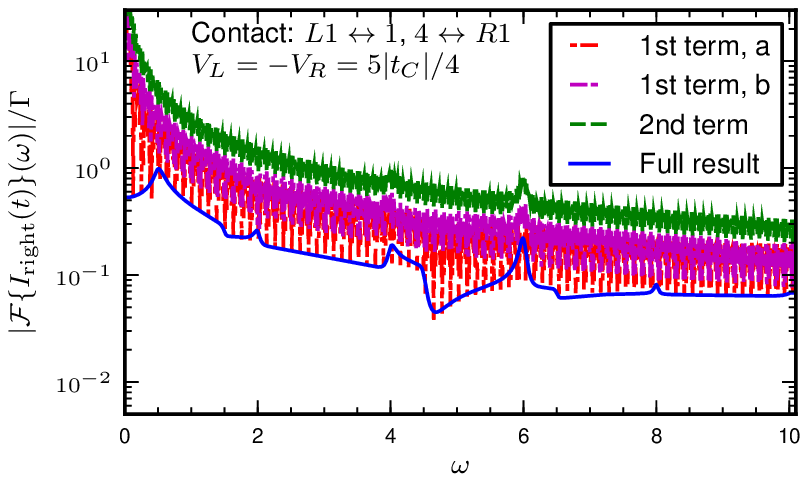}
\caption{\label{fig:res2a}Absolute value of the Fourier transform of 
the terms of the current in units of $\G$. \\}
\end{minipage}
\end{figure}

If we increase the bias window to cover the first molecular levels,  
$V=2.5$, then we see in Fig.~\ref{fig:res1a} that the current has a non-zero 
steady-state value. Similar findings, as with weaker bias, for the possible transitions 
are seen in Fig.~\ref{fig:res2a}. We also see that there is a 
small bump at $\w=4$ in '1st term, b' and '2nd term'. Given that the setup is completely 
identical to the previous case, this fact 
is due to a second (or higher) order response since the same symmetry arguments  
apply.

\begin{figure}[h]
\begin{minipage}{0.45\textwidth}
\centering
\includegraphics[width=\textwidth]{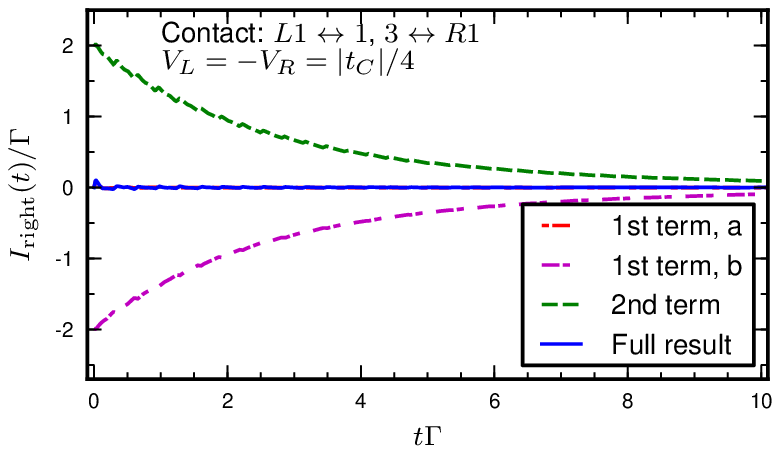}
\caption{\label{fig:res3}Different terms of the time-dependent current in units of $\G$ through
                         the right interface with asymmetric coupling and weak bias.}
\end{minipage}\hspace{2pc}%
\begin{minipage}{0.45\textwidth}
\centering
\includegraphics[width=\textwidth]{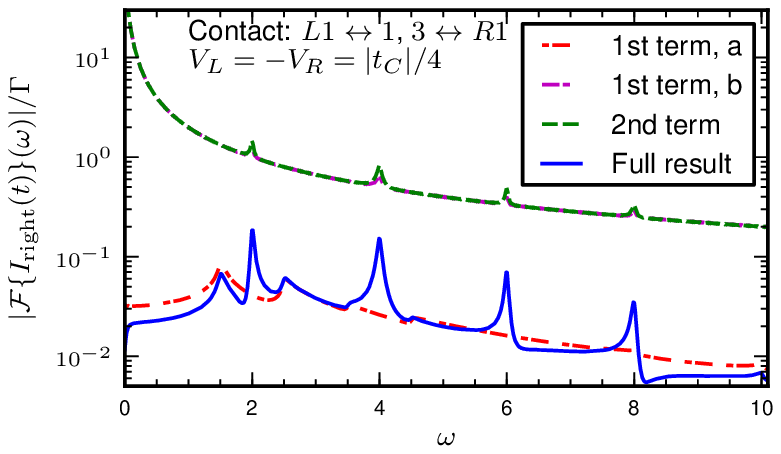}
\caption{\label{fig:res4}Absolute value of the Fourier transform of 
the terms  of the current  in units of $\G$. \\}
\end{minipage}
\end{figure}

\begin{figure}[h]
\begin{minipage}{0.45\textwidth}
\centering
\includegraphics[width=\textwidth]{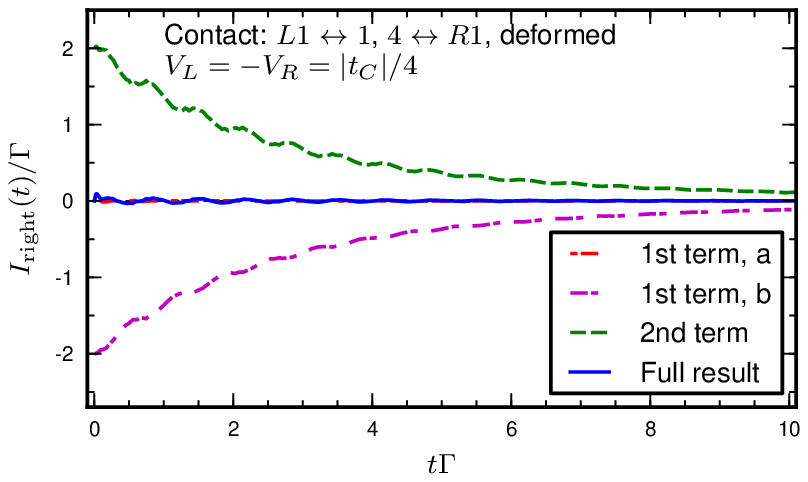}
\caption{\label{fig:res5}Different components of the time-dependent current in units of $\G$ through
                         the right interface of a deformed molecule with
                         symmetric coupling and weak bias.}
\end{minipage}\hspace{2pc}%
\begin{minipage}{0.45\textwidth}
\centering
\includegraphics[width=\textwidth]{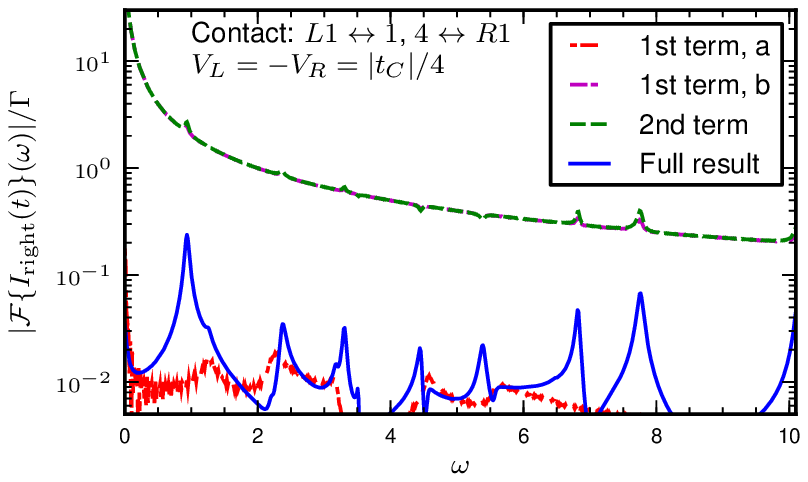}
\caption{\label{fig:res6}Absolute value of the Fourier transform of 
the terms  of the current  in units of $\G$.\\}
\end{minipage}
\end{figure}

If we, however, distort the symmetry of the junction then also
the intramolecular transitions with lower energies become visible. This
is clearly seen in Figs.~\ref{fig:res3} and~\ref{fig:res4} where we connect the
molecule asymmetrically to the leads (1st site to the left and 3rd site to the right, see 
Fig.~\ref{fig:schematic-sites}).
We can also break the symmetry by deforming the molecule with, for instance, one
hopping (between sites $1$ and $6$) being $2t_C$.
This splits the degenerate levels in Fig.~\ref{fig:schematic-levels} and also the
corresponding intramolecular transitions can be seen in Figs.~\ref{fig:res5}
and~\ref{fig:res6}.

\begin{figure}[h]
\begin{minipage}{0.45\textwidth}
\centering
\includegraphics[width=\textwidth]{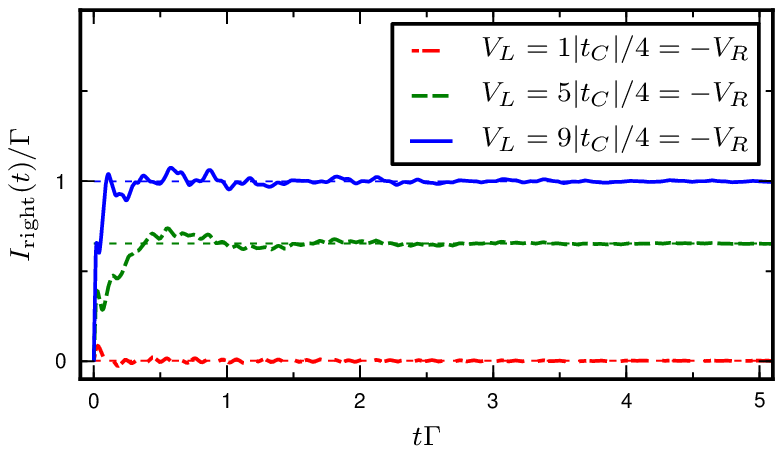}
\caption{\label{fig:res7}Time-dependent current in units of $\G$ through
                         the right interface with symmetric coupling
          	             ($\G=0.1$) and varying bias. (Dotted lines refer to steady-state values.)}
\end{minipage}\hspace{2pc}%
\begin{minipage}{0.45\textwidth}
\centering
\includegraphics[width=\textwidth]{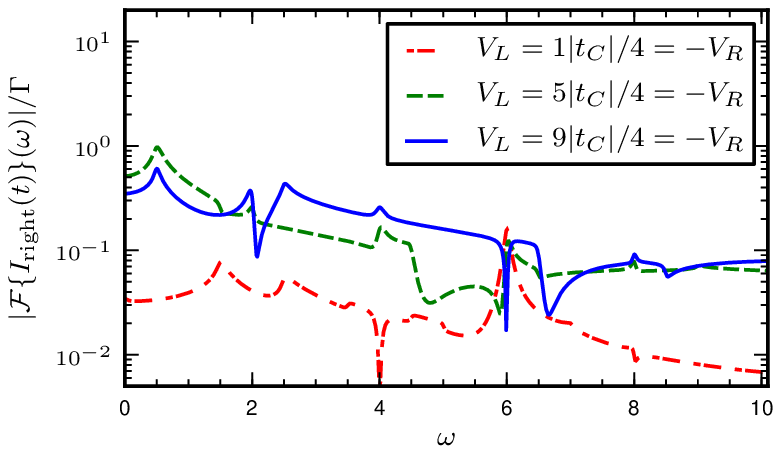}
\caption{\label{fig:res7a}Absolute value of the Fourier transformed right current
                         in units of $\G$. \textcolor{white}{white background text to level
                         the figures. have some more white text. have some more white text.}}
\end{minipage}
\begin{minipage}{0.45\textwidth}
\centering
\includegraphics[width=\textwidth]{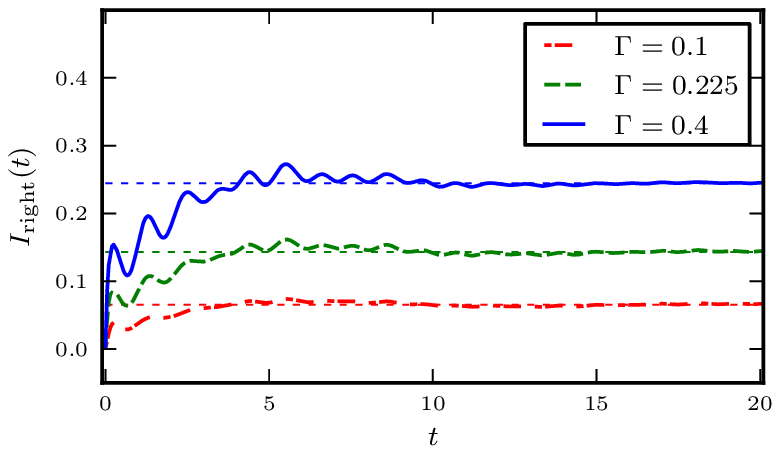}
\caption{\label{fig:res8}Time-dependent current through
                         the right interface with symmetric coupling
                         (varying strength) and bias $V_L = -V_R = 2.5$. (Dotted lines refer to steady-state values.)}
\end{minipage}\hspace{2pc}%
\begin{minipage}{0.45\textwidth}
\centering
\includegraphics[width=\textwidth]{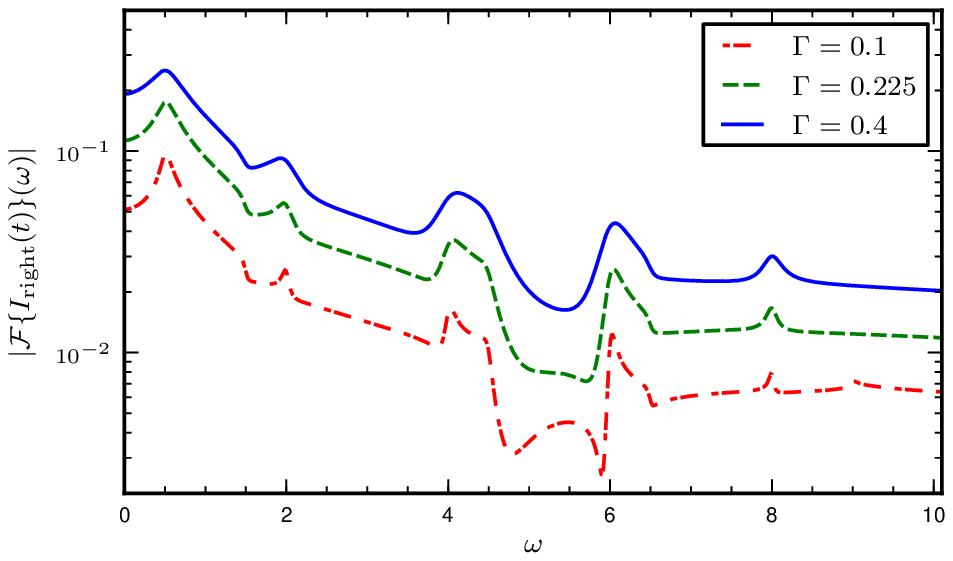}
\caption{\label{fig:res8a}Absolute value of the Fourier transformed right current.
                         \textcolor{white}{white background text to level
                         the figures. have some more white text. have some more white text.} }
\end{minipage}
\end{figure}

As the contributions from different terms sum up to the total current
we can plot the full results for, e.g., the right current of the symmetrically coupled
molecule against, e.g., the bias or the coupling strength. 
In Figs.~\ref{fig:res7}, \ref{fig:res7a}, \ref{fig:res8} 
and~\ref{fig:res8a} we display
the full results. The transient dynamics is visualized better but 
distinguishing between the different contributions is more complicated. 
In Fig. \ref{fig:res8} 
and~\ref{fig:res8a} the axes are not scaled due to varying $\Gamma$.
It is clear that by increasing the bias window more levels open up for transport
 and therefore the steady-state current grows. The oscillation frequencies 
corresponding to transitions between molecular levels remain unchanged while the oscillation
frequencies corresponding to transitions between the molecule and the leads vary (peak shift).
By increasing $\G$, and hence by widening the resonances, electrons 
can flow even with intermediate bias voltages. Correspondingly, the 
steady-state value of the current increases, the relaxation 
time decreases whereas  the oscillation frequencies remain invariant.

\section{Conclusions}

In conclusion we  solved the Kadanoff--Baym equations for the Green's 
function of an open noninteracting system by properly taking into 
account the initial contacts between the system and the reservoirs. 
We used the analytic solution for the time-dependent density 
matrix to derive a time-dependent version of the Landauer--Büttiker 
formula. As an application we considered a tight-binding 
benzene-shaped junction and calculated the time-dependent current 
flowing through it. The advantages of having an explicit solution are 
that the numerical effort is drastically reduced and that the transient 
behavior can easily be interpreted in terms of virtual transitions 
and decay rates.
Our time-dependent Landauer--Büttiker formula holds 
promise for studying the transient behavior of large junctions like, e.g., 
wide nanoribbons or large-diameter nanotubes, as well as disordered junctions 
where a large number of simulations is required to perform the 
average over different configurations.

\ack
R.T. wishes to thank Ellen and Artturi Nyyssönen's foundation for
financial support and CSC --- the Finnish IT Center for Science ---
for providing computing resources. We also acknowledge Petri
Myöhänen, Anna-Maija Uimonen, Niko Säkkinen and Markku Hyrkäs for
productive discussions.

\appendix

\section{Self-energy calculations}
\label{sec:app-self}

The Matsubara self-energy is an antiperiodic function (we are studying
fermions) with period given by the inverse temperature $\beta$.
For the calculation of the Fourier coefficients we can use 
the relation: $\SM_{\a}(\w\pm\im\eta) =
\itS_{\a}^{\text{R}/\text{A}}(\w+\mu)$. Therefore the
Matsubara self-energy is simply given by
\beq\label{eq:self-mats}
\SM_{\a,mn}(\t,\t') & = &
\frac{1}{-\im\b}\sum_{q}\ex^{-\w_q(\t-\t')}\SM_{\a,mn}(\w_q)
\eeq
where
\be
\SM_{\a,mn}(\w_q)=\left\{\begin{array}{ll}
-\frac{\im}{2}\itG_\a & \Im[\w_q] > 0
\\ \\+\frac{\im}{2}\itG_\a & \Im[\w_q] < 0
\end{array}\right.
\ee
and $\w_q = (2q+1)\pi/(-\im\b)$ are the Matsubara frequencies. For
the isolated Green's function of the biased $\a$:th reservoir we have
\beq
\gR_{k\a}(t,t_0) & = &
-\im\theta(t-t_0)\ex^{-\im(\eps_{k\a}+\Val)(t-t_0)} \ , \\
\gM_{k\a}(\t,\t') & = &
\frac{1}{-\im\b}\sum_{q}\frac{\ex^{-\w_q(\t-\t')}}{\w_q-\eps_{k\a}+\mu}
\ .
\eeq
Without loss of generality we take the time $t_{0}$ at which the bias 
is switched on to be zero. Then we can write
\beq
g_{k\a}^{\rceil}(t,\t) & = & \im\gR_{k\a}(t,0)\gM_{k\a}(0,\t) =
\ex^{-\im(\eps_{k\a}+\Val)t}\gM_{k\a}(0,\t) \ ,
\label{eq:isol-g-right} \\
g_{k\a}^{\lceil}(\t,t) & = & -\im\gM_{k\a}(\t,0)\gA_{k\a}(0,t) =
\ex^{\im(\eps_{k\a}+\Val)t}\gM_{k\a}(\t,0) \ . \label{eq:isol-g-left}
\eeq
By using Eqs.~\eqref{eq:isol-g-right} and~\eqref{eq:isol-g-left} we
can calculate the right and left embedding self-energies 
\beq
\itS_{\a,mn}^{\rceil}(t,\t) & = &
\frac{1}{-\im\b}\sum_{q}\ex^{\w_q\t}\sum_{k}T_{m,k\a}\frac{\ex^{-\im(\eps_{k\a}+\Val)t}}{\w_q-\eps_{k\a}+\mu}T_{k\a,n}
\nonumber \\
& = &
\itG_{\a,mn}\frac{1}{-\im\b}\sum_{q}\ex^{\w_q\t}\intw\frac{\ex^{-\im(\w+\Val)t}}{\w_q-\w+\mu}
\ , \label{eq:self-right} \\
\itS_{\a,mn}^{\lceil}(\t,t) & = &
\frac{1}{-\im\b}\sum_{q}\ex^{-\w_q\t}\sum_{k}T_{m,k\a}\frac{\ex^{\im(\eps_{k\a}+\Val)t}}{\w_q-\eps_{k\a}+\mu}T_{k\a,n}
\nonumber \\
& = &
\itG_{\a,mn}\frac{1}{-\im\b}\sum_{q}\ex^{-\w_q\t}\intw\frac{\ex^{\im(\w+\Val)t}}{\w_q-\w+\mu}
\ , \label{eq:self-left}
\eeq
where we used 
$\itG_{\a,mn}=2\pi\sum_{k}T_{mk\a}\d(\w-\eps_{k\a}-\Val)T_{k\a n}$.
It only remains to calculate the \emph{lesser} component. 
We have
\be
g^<_{k\a}(t,t') = \im
f(\eps_{k\a}-\mu)\ex^{-\im(\eps_{k\a}+\Val)(t-t')}
\ee
and therefore
\beq\label{eq:self-lss}
\itS_{\a,mn}^<(t,t') & = & \sum_{k}T_{m,k\a}\im
f(\eps_{k\a}-\mu)\ex^{-\im(\eps_{k\a}+\Val)(t-t')}T_{k\a,n}\nonumber
\\
& = & \im\itG_{\a,mn}\intwf\ex^{-\im(\w+\Val)(t-t')} \, .
\eeq

\section{Green's function calculations}\label{sec:app-green}

The Fourier coefficient of the Matsubara Green's function reads
\beq\label{eq:green-mats}
\GM(\w_q) & = & \frac{1}{\w_q-h-\SM_{\rm em}(\w_q)+\mu} =
\begin{cases}\frac{1}{\w_q-h+\frac{\im}{2}\itG+\mu} & \Im[\w_q] > 0
\\ \frac{1}{\w_q-h-\frac{\im}{2}\itG+\mu} & \Im[\w_q] < 0 \end{cases}
\nonumber \\
& = & \begin{cases}\frac{1}{\w_q-\heff+\mu} & \Im[\w_q] > 0 \\
\frac{1}{\w_q-\heffd+\mu} & \Im[\w_q] < 0 \ , \end{cases}
\eeq
where we defined $\heff = h-\frac{\im}{2}\itG$. The \emph{right}
component of the Green's function can be derived from the equation of
motion
\be\label{eq:green-right-eom}
\left[\idt-h\right]G^{\rceil}(t,\t) =
\int_0^\infty\ud\tb\SR_{\rm em}(t,\tb)G^{\rceil}(\tb,\t)-\im\int_0^\b\ud\tgb\itS_{\rm em}^{\rceil}(t,\tgb)\GM(\tgb,\t)
\ .
\ee
The insertion of the retarded self-energy from Eq.~\eqref{eq:wbla} leads to
\be\label{eq:green-right-deriv}
G^\rceil(t,\t) = \ex^{-\im\heff t}\Big[\GM(0,\t)-\int_0^t\ud
t'\ex^{\im\heff
t'}\int_0^\beta\ud\tgb\itS_{\rm em}^\rceil(t',\tgb)\GM(\tgb,\t)\Big] \ ,
\ee
where we noticed that $\GM(0,\t) = G^\rceil(0,\t)$. 

Finally the retarded Green's function in Fourier space reads
\be
\label{eq:green-ret-dyson}
\GR(\w) =\frac{1}{\w-h-\SR_{\rm em}(\w)} =\frac{1}{\w-h+\frac{\im}{2}\itG} 
\ee
and Fourier transforming back in the time domain we recover Eq. 
(\ref{gret}).

\section{The three terms in Eq.~\eqref{eq:final-glss}}
\label{sec:app-terms}
 For the first term we use Eqs.~\eqref{gret}
and~\eqref{eq:self-lss} to obtain
\beq
\left[\GR\cdot\itS_{\rm em}^<\right](t,t) & = &
\int_0^\infty\ud\tb\,\GR(t,\tb)\itS_{\rm em}^<(\tb,t) \nonumber \\
& = &
\im\sum_\a\intwf\left[1-\ex^{\im(\w+\Val-\heff)}\right]\GR(\w+\Val)\itG_\a
\ .
\eeq
Also the second term is readily calculated by using
Eq.~\eqref{eq:self-wbla} \be
\left[G^<\cdot\SA_{\rm em}\right](t,t) = \int_0^\infty\ud\tb\,
G^<(t,\tb)\SA_{\rm em}(\tb,t) = \frac{\im}{2}G^<(t,t)\itG \ .
\ee
The third term involves somewhat more trickery because of the rather
complicated form of the right Green's function. Inserting the
expressions from Eqs.~\eqref{eq:green-right-deriv}
and~\eqref{eq:self-left} we get
\beq\label{eq:third-term-deriv1}
\left[G^\rceil\star\itS_{\rm em}^\lceil\right](t,t) & = & 
-\im\int_0^\b\ud\t\,
G^\rceil(t,\t)\itS_{\rm em}^\lceil(\t,t) \nonumber \\
& = & \ex^{-\im\heff t}\left\{\left[\GM\star\itS_{\rm em}^\lceil\right](0,t) -
\im\int_0^t\ud t' \ex^{\im\heff
t'}\left[\itS_{\rm em}^\rceil\star\GM\star\itS_{\rm 
em}^\lceil\right](t',t)\right\} .
\nonumber \\
\eeq
By using $\int_0^\b\ud\t\ex^{(\w_q-\w_{q'})\t}=\b\del{q}{q'}$ for the
Matsubara frequencies we may manipulate  Eq.~\eqref{eq:third-term-deriv1}
further. Inserting  Eqs.~\eqref{eq:self-right},
\eqref{eq:self-left} and~\eqref{eq:green-mats} we obtain for the
double convolution
\beq\label{eq:third-term-deriv2}
\left[\itS_{\rm em}^\rceil\star\GM\star\itS_{\rm em}^\lceil\right](t',t) & = &
-\im\int_0^\b\ud\t
(-\im)\int_0^\b\ud\tgb\itS_{\rm em}^\rceil(t',\tgb)\GM(\tgb,\t)\itS_{\rm em}^\lceil(\t,t)
\nonumber \\
& = &
\intw\intwp\sum_{\a,\a'}\itG_\a\frac{1}{-\im\b}\sum_q
\frac{\ex^{-\im(\w+\Val)t'}}{\w_q-\w+\mu}\GM(\w_q)\frac{\ex^{\im(\w'+\Valp)t}}{\w_q-\w'+\mu}\itG_{\a'}
\ . \nonumber \\
\eeq
The integration with respect to $\w$ can be done by closing the 
contour in lower-half
plane (LHP) because of the exponential convergence factor, whereas
the integration with respect to $\w'$ can be done by closing the 
contour in the upper-half
plane (UHP). However, the corresponding poles are located on
different half planes, and this makes the double integral to vanish
for every $\w_q$. Hence
\be\label{eq:third-term-deriv3}
\left[\Sigmaem^\rceil\star\GM\star\Sigmaem^\lceil\right](t',t) = 0 \ ,
\ee
and in Eq.~\eqref{eq:third-term-deriv1} we are left with
\beq\label{eq:third-term-deriv4}
\left[\GM\star\Sigmaem^\lceil\right](0,t) & = &
-\im\int_0^\b\ud\t\GM(0,\t)\Sigmaem^\lceil(\t,t) \nonumber \\
& = &
\intw\frac{1}{-\im\b}\sum_q\frac{\GM(\w_q)\ex^{\eta\w_q}}{\w_q-\w+\mu}\sum_\a\itG_\a\ex^{\im(\w+\Val)t}
\ ,
\eeq
where on the last line a convergence factor
$\ex^{\eta\w_q}$ was added to account for correct limiting behaviour
when $t\to 0$. The sum over 
Matsubara frequencies can be performed using the Luttinger--Ward trick 
\cite{lwpaper} and yields
\beq\label{eq:third-term-deriv6}
\frac{1}{-\im\b}\sum_q\frac{\GM(\w_q)\ex^{\eta\w_q}}{\w_q-\w+\mu} & =
&
\int_\g\frac{\ud\w'}{2\pi}f(\w')\ex^{\eta\w'}\frac{\GM(\w')}{\w'-\w+\mu}
\nonumber \\
& = &
\int_{-\infty}^\infty\frac{\ud\w'}{2\pi}f(\w')
\left[\frac{\GM(\w'-\im\delta)}{\w'-\w+\mu-\im\delta}-
\frac{\GM(\w'+\im\delta)}{\w'-\w+\mu+\im\delta}\right]
\ .
\eeq
By inserting Eq.~\eqref{eq:third-term-deriv6} into
Eq.~\eqref{eq:third-term-deriv4} we get
\be\label{eq:third-term-deriv7}
\left[\GM\star\itS_{\rm em}^\lceil\right](0,t) = \intw\intwp
f(\w')\left[\frac{\GM(\w'-\im\delta)}{\w'-\w+\mu-\im\delta}-\frac{\GM(\w'+\im\delta)}{\w'-\w+\mu+\im\delta}\right]\sum_\a\itG_\a\ex^{\im(\w+\Val)t}
\ .
\ee
Now the integral over $\w$ can be done by closing the contour in the 
UHP. The first term in square brackets
integrates to zero because of the pole in the LHP. The pole of the
second term occurs at $\w =
\w'+\mu+\im\delta$, and therefore
\beq\label{eq:third-term-deriv8}
\left[\GM\star\itS_{\rm em}^\lceil\right](0,t) & = & \im\intwp
f(\w')\GM(\w'+\im\delta)\sum_\a\ex^{\im(\w'+\mu+\im\delta+\Val)t}
\nonumber \\
\begin{scriptsize}\left.\begin{matrix}\delta \to 0^+ \\ \w' =
\w-\mu\end{matrix} \ \right\} \ \rightarrow\end{scriptsize} \ & = &
\im\intwf\underbrace{\GM(\w-\mu)}_{=\GR(\w)}\sum_\a\itG_\a\ex^{\im(\w-\mu+\mu+\Val)t}
\nonumber \\
& = & \im\intwf\GR(\w)\sum_\a\itG_\a\ex^{\im(\w+\Val)} \ .
\eeq
This can be inserted into Eq.~\eqref{eq:third-term-deriv1} to obtain
\be
\left[G^\rceil\star\itS_{\rm em}^\lceil\right](t,t) =
\im\intwf\sum_\a\ex^{\im(\w+\Val-\heff)t}\GR(\w)\itG_\a \ .
\ee

\section{Derivation of Eq.~\eqref{eq:glss-final2}}
\label{sec:app-manip}

We first state some useful identities for the retarded/advanced Green's
function to be used later:
\be\label{eq:green-ret-identity}
G^{\rm R/A}(\w+\Val) = G^{\rm R/A}(\w)-\Val G^{\rm R/A}(\w)G^{\rm R/A}(\w+\Val) \ ,
\ee
which can be checked directly by using Eq. (\ref{eq:green-ret-dyson}) 
and its adjoint. From Eq. (\ref{eq:green-ret-identity}) it follows 
that
\be\label{eq:green-ret-adv-identity}
\Val^2\GR(\w)\GA(\w)\GR(\w+\Val)\GA(\w+\Val) =
\left[\GR(\w)-\GR(\w+\Val)\right]\left[\GA(\w)-\GA(\w+\Val)\right] \ .
\ee
Taking into account Eq.~\eqref{eq:green-ret-identity} in
Eq.~\eqref{eq:glss-deriv1} we get
\beq\label{eq:glss-deriv11}
& & \idt G^<(t,t) - \heff G^<(t,t)+G^<(t,t)\heffd \nonumber \\
& = &
-\im\intwf\sum_\a\left\{\ex^{\im(\w+\Val-\heff)t}\Val\GR(\w)\GR(\w+\Val)+\GR(\w+\Val)\right\}\itG_\a
\nonumber \\
& + &
\im\intwf\sum_\a\itG_\a\left\{\GA(\w+\Val)\GA(\w)\Val\ex^{-\im(\w+\Val-\heffd)t}+\GA(\w+\Val)\right\}
\ .
\eeq
It is convenient to rewrite the Green's function as $G^<(t,t) =
\ex^{-\im\heff t}\tilde{G}^<(t,t)\ex^{\im\heffd t}$. In this way the
left-hand side of Eq.~\eqref{eq:glss-deriv11} becomes
\beq\label{eq:glss-deriv2}
 & & \idt\left[\ex^{-\im\heff t}\tilde{G}^<(t,t)\ex^{\im\heffd
t}\right]-\heff\ex^{-\im\heff t}\tilde{G}^<(t,t)\ex^{\im\heffd
t}+\ex^{-\im\heff t}\tilde{G}^<(t,t)\ex^{\im\heffd t}\heffd \nonumber
\\
& = & \ex^{-\im\heff t}\idt \tilde{G}^<(t,t)\ex^{\im\heffd t} \ .
\eeq
Then the right-hand side of Eq.~\eqref{eq:glss-deriv11} can be
multiplied from left by $\ex^{\im\heff t}$ and from right by
$\ex^{-\im\heffd t}$ to give
\beq\label{eq:glss-deriv3}
\idt \tilde{G}^<(t,t) & = & -\im\intwf\sum_\a\ex^{\im\heff
t}\left[\GR(\w+\Val)\itG_\a-\itG_\a\GA(\w+\Val)\right]\ex^{\im\heffd
t} \nonumber \\
& &
-\im\intwf\sum_\a\Val\left[\GR(\w)\GR(\w+\Val)\itG_\a\ex^{\im(\w+\Val-\heffd)t}\right.
\nonumber \\
& & \hspace{120pt}\left.
-\ex^{-\im(\w+\Val-\heff)t}\itG_\a\GA(\w+\Val)\GA(\w)\right] \ .
\eeq
Now we are ready to integrate both sides over $t$ to obtain
\beq\label{eq:glss-deriv4}
\tilde{G}^<(t,t)-\underbrace{\tilde{G}^<(0,0^+)}_{\mathclap{=G^<(0,0^+)
= \GM(0,0^+)}} & = & -\intwf\sum_\a\int_{0}^{t}\ud t'
\nonumber \\
& & \times
\left\{\ex^{\im\heff
t'}\left[\GR(\w+\Val)\itG_\a-\itG_\a\GA(\w+\Val)\right]\ex^{-\im\heffd
t'} \right.\nonumber \\
& & \hspace{0.5cm}-\Val\left[\GR(\w)\GR(\w+\Val)\itG_\a\ex^{\im(\w+\Val-\heffd)t'}
\right.\nonumber \\
& & \left.\left.\hspace{1.5cm} -
\ex^{-\im(\w+\Val-\heff)t'}\itG_\a\GA(\w+\Val)\GA(\w)\right] \right\}\ .
\eeq
The integration over $t'$ for the second term in
Eq~\eqref{eq:glss-deriv4} can easily be done. For the first
term we need the following result: Given two arbitrary matrices $A$ and $B$
\be\label{eq:glss-deriv6}
\int_0^t\ud t'\ex^{\im A
t'}\left[\frac{1}{x-A}B-B\frac{1}{x-A^\dagger}\right]\ex^{-\im
A^\dagger t'} = -\im\ex^{\im A
t'}\frac{1}{x-A}B\frac{1}{x-A^\dagger}\ex^{-\im A^\dagger t'} \ ,
\ee
which can directly be verified by differentiating the right-hand side
with respect to $t'$. Applying this result to
Eq.~\eqref{eq:glss-deriv4} we obtain
\beq\label{eq:glss-deriv7}
\tilde{G}^<(t,t) & = & \im\intwf\sum_\a\left\{\GR(\w)\itG_\a\GA(\w) +
\ex^{\im\heff t}\GR(\w+\Val)\itG_\a\GA(\w+\Val)\ex^{-\im\heffd
t}\right. \nonumber \\
& & \left.\hspace{85pt}-\GR(\w+\Val)\itG_\a\GA(\w+\Val)
\right.\nonumber \\ \nonumber \\
& & \left.\hspace{85pt}
+\Val\GR(\w)\GR(\w+\Val)\itG_\a\GA(\w)\GA(\w+\Val)\ex^{\im(\w+\Val-\heffd)t}\right.
\nonumber \\ \nonumber \\
& & \left.\hspace{85pt}
-\Val\GR(\w)\GR(\w+\Val)\itG_\a\GA(\w)\GA(\w+\Val) \right.\nonumber
\\ \nonumber \\
& & \left.\hspace{85pt} +
\Val\GR(\w+\Val)\ex^{-\im(\w+\Val-\heff)t}\itG_\a\GA(\w+\Val)\GA(\w)
\right. \nonumber \\ \nonumber \\
& & \left.\hspace{85pt}
-\Val\GR(\w+\Val)\itG_\a\GA(\w+\Val)\GA(\w)\right\} \ .
\eeq
Then the definition for $\tilde{G}$ can be inserted into the
left-hand side, and multiply accordingly with $\ex^{-\im\heff t}$
from left and with $\ex^{\im\heffd t}$ from right. 
Combining terms according to Eqs.~\eqref{eq:green-ret-identity}
and~\eqref{eq:green-ret-adv-identity} we find Eq. 
(\ref{eq:glss-final2}).

\bibliography{iopart-num}

\end{document}